\documentstyle[12pt,aasms4]{article}

\lefthead{Hodge et al}
\righthead{WLM Globular Cluster}

\begin{document}

\title{HST Studies of the WLM Galaxy. I. The Age and Metallicity of the
Globular Cluster
\footnote{
Based on observations with the NASA/ESA Hubble Space Telescope obtained
at the Space Telescope Science Institute, which is operated by the
Association of Universities for Research in Astronomy, Inc., under
NASA contract NAS5-26555.
}}

\author{Paul W.\ Hodge, Andrew E.\ Dolphin, and Toby R.\ Smith}
\affil{Department of Astronomy, University of Washington, Box 351580,
Seattle, WA 98195-1580 \\
hodge@astro.washington.edu, dolphin@astro.washington.edu,
smith@astro.washington.edu}

\and

\author{Mario Mateo}
\affil{Astronomy Department, University of Michigan, Ann Arbor, MI 48109-1090\\
mateo@astro.lsa.umich.edu}

\begin{abstract}
We have obtained $V$ and $I$ images of the lone globular cluster that belongs
to the dwarf Local Group irregular galaxy known as WLM. The color-magnitude
diagram
of the cluster shows that it is a normal old globular cluster with a
well-defined giant branch reaching to $M_V=-2.5$, a horizontal branch at
$M_V=+0.5$, and a sub-giant branch extending to our photometry limit of 
$M_V=+2.0$.  A best fit to theoretical isochrones indicates that this cluster
has a metallicity of [Fe/H]$=-1.52\pm0.08$ and an age of $14.8\pm0.6$ Gyr,
thus indicating that it is similar to normal old halo globulars in our
Galaxy. From the fit we also find that the distance modulus of the cluster
is $24.73\pm0.07$ and the extinction is $A_V=0.07\pm0.06$, both values
that agree within the errors with data obtained for the galaxy itself by
others. We conclude that this normal massive cluster was able to form
during the formation of WLM, despite the parent galaxy's very small
intrinsic mass and size.
\end{abstract}

\keywords{galaxies: star clusters --- Local Group}

\section{Introduction}

The galaxy known as WLM is a low-luminosity, dwarf irregular galaxy in the
Local Group. A history of its discovery and early study was given by
Sandage \& Carlson (1985). Photographic surface photometry of the galaxy
was published by Ables \& Ables (1977). Its stellar population has been
investigated from ground-based observations by Ferraro et al (1989) and by
Minniti \& Zijlstra (1997). The former showed that the main body of the
galaxy consists of a young population, which dominates the light, while the
latter added the fact that there appears to be a very old population in
its faint outer regions.  Cepheid variables were detected by Sandage \&
Carlson (1985), who derived its distance, and were reanalyzed by Feast \&
Walker
(1987) and by Lee et al. (1992). The latter paper used $I$ photometry of
the Cepheids and the RGB (red giant branch) distance criterion to conclude
that the distance modulus for WLM is 24.87 $\pm$ 0.08. The extinction
determined by Feast \& Walker (1987) is $A_B$ = 0.1.

Humason et al. (1956), when measuring the radial velocity of WLM, noticed a
bright
object next to it that had the appearance of a globular cluster. Its radial
velocity was the same as that of WLM, indicating membership. Ables \&
Ables (1977) found that the cluster's colors were like those of a globular
cluster, and Sandage \& Carlson (1985) confirmed this. Its total
luminosity is unusual for its being the sole globular of a galaxy. Sandage
\& Carlson (1985) quote a magnitude of $V$ = 16.06, indicating an absolute
magnitude of $M_V$ = -8.8.
This can be compared to the mean absolute magnitude of globulars in
galaxies, which is $M_V = -7.1 \pm 0.43$ (Harris 1991). The cluster, though
unusually bright, has only a small fraction of the $V$ luminosity of the
galaxy, which is 5.2 magnitudes brighter in $V$.

One could ask the question of whether there are other massive clusters in
the galaxy, such as luminous blue clusters similar to those in the
Magellanic Clouds. Minniti \& Zijlstra (1997), using the NTT and thus
having a wider field than ours, searched for other globular clusters and
found none. However, the central area of the galaxy has one very young,
luminous cluster, designated C3 in Hodge, Skelton and Ashizawa (1999). This
object is the nuclear cluster of one of the brightest HII regions (Hodge
\& Miller 1995). There do not appear to be any large intermediate age
clusters, such as those in the Magellanic Clouds or that recently
identified spectroscopically in the irregular galaxy NGC 6822 by Cohen \&
Blakeslee (1998) .

No other Local Group irregular galaxy fainter than $M_V$ = -16 contains a
globular cluster. The elliptical dwarf galaxies NGC 147 and NGC 185 (0.8
and 1.3 absolute magnitudes brighter than WLM, respectively) do have a few
globular clusters each and Fornax (1.7 absolute magnitudes fainter) has
five, which makes it quite anomalous, even for an elliptical galaxy (see
Harris, 1991, for references).

Another comparison can be made using the specific frequency parameter, as
defined and discussed by Harris (1991). The value of the specific frequency
calculated for WLM is 7.4, which can be compared to Harris' value
calculated for late-type galaxies, which is $0.5 \pm 0.2$. The highest
average specific frequency is found for nucleated dwarf elliptical galaxies
by Miller et al (1998), which is $6.5 \pm 1.2$, while non-nucleated
dwarf elliptical galaxies have an average of $3.1 \pm 0.5$.  These values
are similar to those found by Durrell et al (1996), implying that the
specific frequency for WLM is comparable to that for dwarf elliptical
galaxies but possibly higher than that for other late-type galaxies.

Because the WLM cluster as a globular in an irregular dwarf galaxy is
unique, it may represent an unusual opportunity to
investigate the question of whether Local Group dwarf irregulars share the
early history of our Galaxy and other more luminous Group members, which formed
their massive clusters some 15 Gyr ago, or whether they formed later, as
the early ideas about the globular clusters of the Magellanic Clouds seemed
to indicate. Of course, we now know that the LMC has several true globular
clusters that are essentially identical in age to the old halo globulars of
our Galaxy (Olsen et al. 1998, Johnson et al. 1998), so the evidence
suggesting a delayed formation now seems to come only from the SMC. In any
case, WLM gives us a rare opportunity to find the oldest cluster (and
probably the oldest stars) in a more distant and intrinsically much less
luminous star-forming galaxy in the Local Group.

\section{Data and Reduction}

\subsection{Observations}

As part of a Cycle 6 HST GO program, we obtained four images of the WLM
globular cluster on 26 September, 1998.  There were two exposures taken
with the F814W filter of 2700 seconds and two with the F555W filter, one
each of 2700 seconds and 2600 seconds. The globular cluster was centered on
the PC chip and the orientation of the camera was such that the WF chips
lay approximately along the galaxy's minor axis, providing a representative
sample of the WLM field stars to allow us to separate cluster stars
reliably.

\subsection{Reductions}

With two images of equal time per filter, cosmic rays were cleaned with
an algorithm nearly identical to that used by the IRAF task CRREJ.  The
two images were compared at each pixel, with the higher value thrown out
if it exceeded 2.5 sigma of the average.  The cleaned, combined F555W
image is shown in Figure \ref{fig_image}.

\placefigure{fig_image}

Photometry was then carried out using a program specifically designed to
reduce undersampled WFPC2 data.  The first step was to build a library
of synthetic point spread functions (PSFs), for which Tiny Tim 4.0
(Krist 1995) was used.
PSFs were calculated at 49 positions on each chip in F555W and F814W,
subsampled at 20 per pixel in the WF chips and 10 in the PC chip.  The
subsampled PSFs were adjusted for charge diffusion and estimated subpixel
QE variations, and combined for various locations of a star's center
within a pixel.  For example, the library would contain a PSF for the
case of a star centered in the middle of a pixel, as well as for a star
centered on the edge of a pixel.  In all, a 10x10 grid of possible centerings
was made for the WF chips and a 5x5 grid for the PC chip.
This served as the PSF library for the photometry.

The photometry was run with an iterative fit that located stars and
then found the best combinations of stellar profiles to match the
image.  Rather than using a centroid to determine which PSF to use for
a star, a fit was attempted with each PSF centered near the star's position
and the best-fitting PSF was chosen.  This method helped
avoid the problem of centering on an undersampled image.  Residual
cosmic rays and other non-stellar images were removed through a
chi-squared cut over the final photometry list.

The PSF fit was normalized to give the total number of counts from the
star falling within a 0.5 arcsec radius of the center.  This count rate was
then converted into magnitudes as described in Holtzman et al (1995), using
the CTE correction, geometric corrections, and transformation.
For the color-magnitude diagram (CMD) and luminosity
function analyses below, roughly the central 20\% of the image was analyzed
to maximize the signal from the globular while minimizing the background
star contamination.  In that region, the effect of background stars is
negligible.
The CMD from this method, showing all stars observed, is shown in Figure
\ref{fig_phot}a, with the same data reduced with DAOPHOT shown in Figure
\ref{fig_phot}b.  Error bars are shown corresponding to the artificial star
results (which account for crowding in addition to photon statistics),
rather than the standard errors from the PSF fits and
transformations.  The photometry list is given in Table \ref{tabphot},
which will appear in its entirety only in the electronic edition.  The
table contains X and Y positions of each star, with $V$ and $I$ magnitudes
and uncertainties.  Uncertainties given are from the PSF fitting and
from the transformations.

\placetable{tabphot}
\placefigure{fig_phot}

Artificial star tests were made, with each star
added and analyzed one at a time to minimize additional crowding often
caused by the addition of the artificial stars.  The artificial
stars were added to both the combined $V$ and $I$ images, so that a library
of artificial star results for a given position, $V$ magnitude, and color
could be built.  In addition to completeness corrections, these data were
employed in the generation of synthetic CMDs for the determination
of the star formation history of the cluster.

\section{Analysis}

\subsection{Luminosity Function}

The luminosity functions (LFs) of the globular cluster in $V$ and $I$ are
shown in Figure \ref{fig_lf}a and \ref{fig_lf}b, respectively, binned
into 0.5 magnitude bins.  Theoretical luminosity functions are given
as well, from interpolated Padova isochrones (Girardi et al 1996, Fagotto et
al 1994) using the star formation parameters and distance obtained 
in the CMD analysis below.

\placefigure{fig_lf}

The observed and theoretical LFs are in excellent agreement.  The bump in
the observed $V$ LF between magnitudes 25 and 26, and the bump in the
observed $I$ LF starting at magnitude 24.5 are due to the horizontal
branch stars, which cannot be separated from the rest of the CMD
cleanly.  The only other significant deviation, the bump in the $V$ LF
between magnitudes 22.5 and 23, is the clump of stars at the
tip of the RGB, which is also observed in the CMD.  This seems to be a
statistical
fluke, a result of the relatively small number of stars in that part
of the RGB, and is similar to statistical flukes seen in Monte Carlo
simulations.  Thus as far as can be determined, the observed LF agrees
with the theoretical expectations.

\subsection{Color-Magnitude Diagram}

For the CMD analysis, a cleaner CMD was achieved by omitting all stars
with PSF fits worse than a chi-squared value of 3.
The observed $V, V-I$ CMD is shown in Figure \ref{fig_cmd}a, and was
analyzed as described in Dolphin (1997).  Interpolated Padova isochrones
(Girardi et al 1996, Fagotto et al 1994) were used to generate the
synthetic CMDs, with photometric errors and incompleteness simulated
by application of artificial star results to the isochrones.
No assumptions were made regarding the star formation history, metallicity,
distance, or extinction to the cluster, and a fit was attempted with all of
these parameters free.  The best fits that returned single-population
star formation histories were then combined with a weighted average to
determine the best parameters of star formation.  Uncertainties were
derived by taking a standard deviation of the parameters from the fits,
and thus include the fitting errors and uncertainties resulting from
an age-metallicity-distance ``degeneracy.''  Systematic errors due to the
particular choice of evolutionary models are naturally present, but are not
accounted for in the uncertainties.  The following parameters were
obtained:
\begin{itemize}
\item Age: 14.8 $\pm$ 0.6 Gyr
\item Fe/H: -1.52 $\pm$ 0.08
\item Distance modulus: 24.73 $\pm$ 0.07
\item Av: 0.07 $\pm$ 0.06
\end{itemize}

A synthetic CMD constructed from these parameters is shown in Figure
\ref{fig_cmd}b, using the artificial star data to mimic photometric
errors, completeness, and blending in the data.  The poorly reproduced
horizontal branch is a result of the isochrones we used, but the giant
branch was reproduced well, with the proper shape and position.

\placefigure{fig_cmd}

\subsection{Structure}

Profiles were calculated in bins of 10 pixels (0.45 arcsec) in both the
$V$ and $I$ images, and are shown in Figure \ref{fig_prof}, corrected for
incompleteness (both as a function of magnitude and position).  The cutoff
magnitudes of 27 in $V$ and 26 in $I$ were chosen to minimize the corrections
required due to incompleteness.
Additionally, the central bin (0-10 pixels)
was omitted because of extreme crowding problems.  The remaining
bins were fit to King models with a least-squares fit.  The best
parameters for the King models (assuming a
distance modulus of 24.73) are as follows (shown by the solid lines in Figure
\ref{fig_prof}).
\begin{itemize}
\item core radius: 1.09 $\pm$ 0.14 arcsec (4.6 $\pm$ 0.6 pc)
\item tidal radius: 31 $\pm$ 15 arcsec (130 $\pm$ 60 pc)
\item core density: 59 $\pm$ 8 stars/arcsec$^2$ (3.2 $\pm$ 0.4 stars/pc$^2$) $V$, 44 $\pm$ 6 stars/arcsec$^2$ (2.4 $\pm$ 0.3 stars/pc$^2$) $I$
\item background density: 0.77 $\pm$ 0.12 stars/arcsec$^2$ (0.042 $\pm$ 0.007 stars/pc$^2$) $V$, 0.77 $\pm$ 0.12 stars/arcsec$^2$ (0.042 $\pm$ 0.007 stars/pc$^2$) $I$
\end{itemize}
For a distance modulus of 24.87 (Lee et al. 1992), the corresponding sizes
would be 7\% larger.
For comparison, Trager et al. (1993) find that 2/3 of Milky
Way clusters have core radii between approximately 5 and 60 pc.

\placefigure{fig_prof}

\section{Conclusions}

Our analysis shows that the WLM globular cluster is virtually
indistinguishable from a halo globular in our Galaxy. We find that a
formal fit to theoretical isochrones indicates an age of 14.8 $\pm$ 0.6 Gyr,
which agrees with ages currently being measured for Galactic globulars
(e.g., vandenBerg 1998) and a metallicity of [Fe/H] of -1.52 $\pm$ 0.08, a
typical globular cluster value that is similar to that obtained for the
outer field giant stars along the minor axis of WLM by Minniti and Zijlstra
(1997) and by us (Dolphin 1999). The distance modulus for the cluster,
derived independently from the parent galaxy, is 24.73 $\pm$ 0.07, which
agrees within the errors with that derived from Cepheids and the RGB (Lee
et al. 1992) of the galaxy.

In structure the globular is elongated in outline, with a mean radial profile
that fits a King (1962) model within the observational uncertainties. We
derive a core radius of 1.09 $\pm$ 0.14 arcsec and a tidal radius of 31 $\pm$
15 arcsec, which translate to 4.6 $\pm$ 0.6 pc and 130 $\pm$ 60 pc,
respectively. The core radius is very similar to that found for massive
globulars in our Galaxy (Trager et al. 1993), while the tidal radius,
though quite uncertain, is rather large by comparison. The former result
indicates that formation conditions in this galaxy near its conception were
such that a massive, highly concentrated star cluster could form, despite
the very small amount of the total mass of material available. The latter
result is probably an indication that the tidal force of the galaxy on the
cluster is small.

The presence of a normal, massive globular cluster in this dwarf irregular
galaxy may be an useful piece of evidence regarding the early history of
star, cluster and galaxy formation. Recent progress in the field of
globular cluster formation has resulted from both observational and
theoretical studies (Searle \& Zinn 1978, Harris \& Pudritz 1994,
McLaughlin \& Pudritz 1994, Durrell et al 1996, Miller et al 1998, and
McLaughin 1999).  Although the uncertainties from a single data point are
sufficiently large to discourage quantitative analysis, the presence of
a globular cluster in WLM would constrain formation models that predict
$\ll 1$ cluster in such a galaxy.

\acknowledgements

We are indebted to the excellent staff of the Space Telescope Science
Institute for obtaining these data and to NASA for support of the analysis
through grant GO-06813.

\begin{table}
\dummytable\label{tabphot}
\end{table}

\clearpage

\figcaption[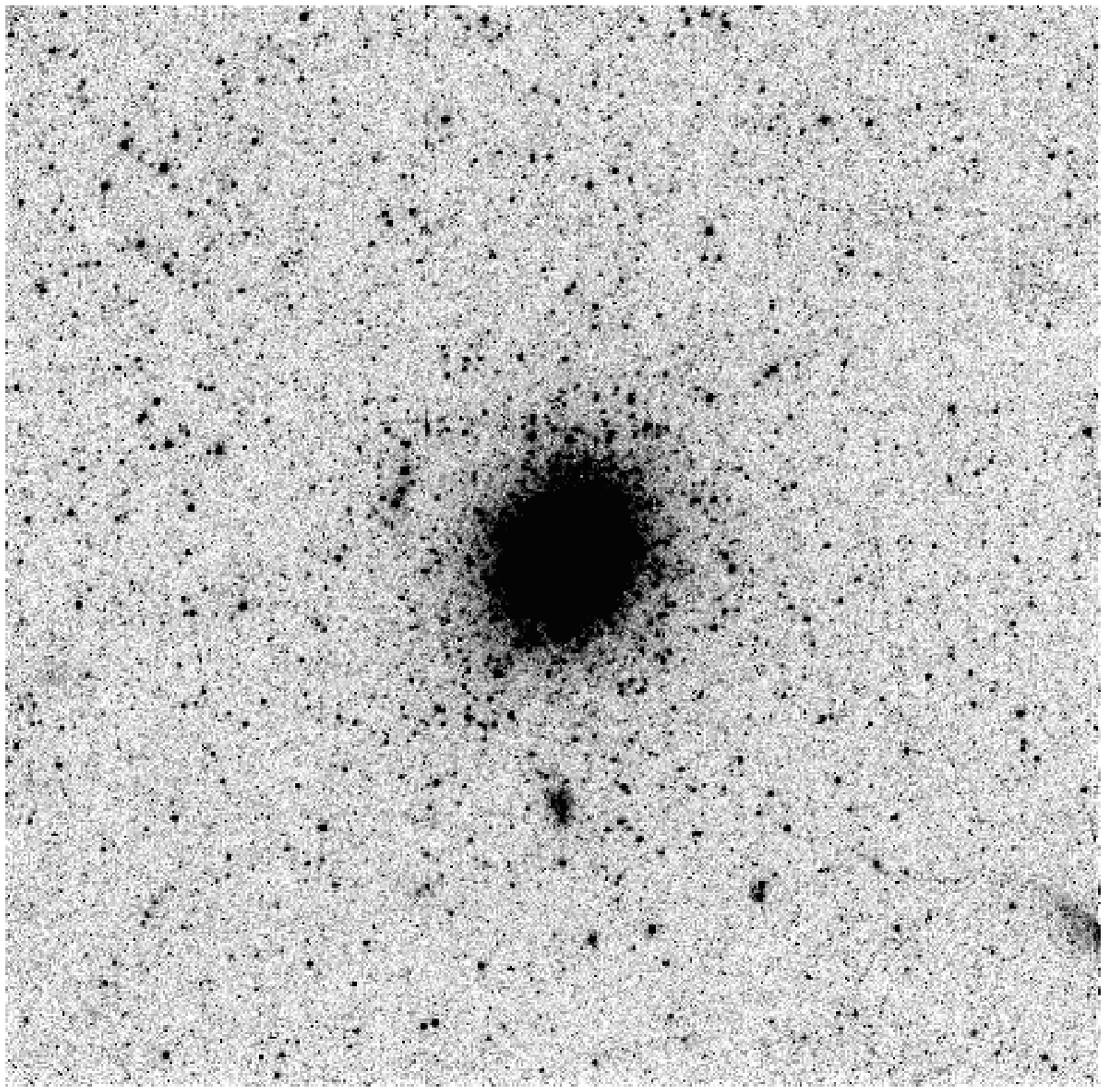]{Combined $V$ image of the WLM globular cluster after cosmic ray cleaning. \label{fig_image}}
\figcaption[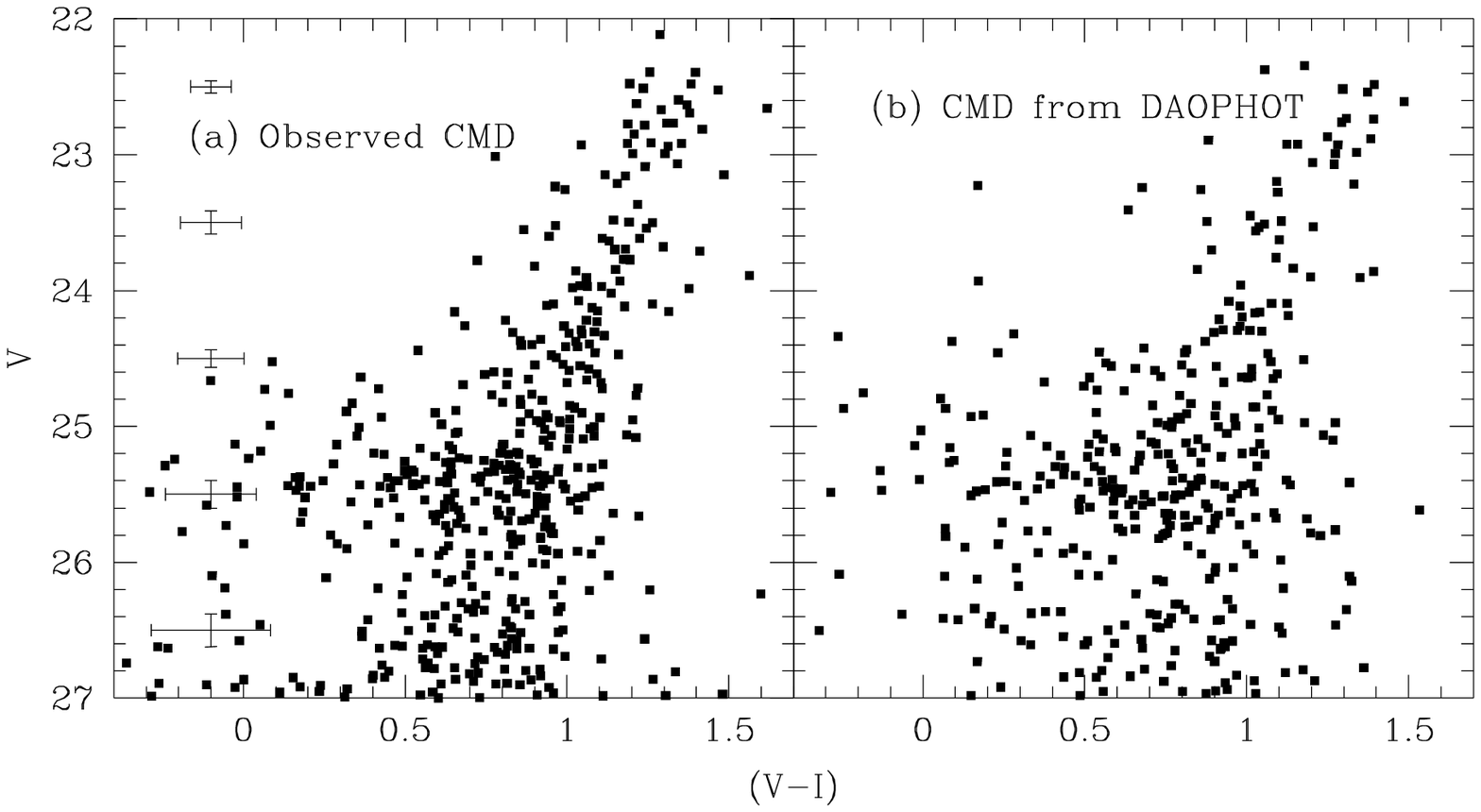]{The $V, V-I$ color-magnitude diagram of the cluster is shown in figure (a).  Figure (b) shows the same data, reduced with the IRAF task DAOPHOT. \label{fig_phot}}
\figcaption[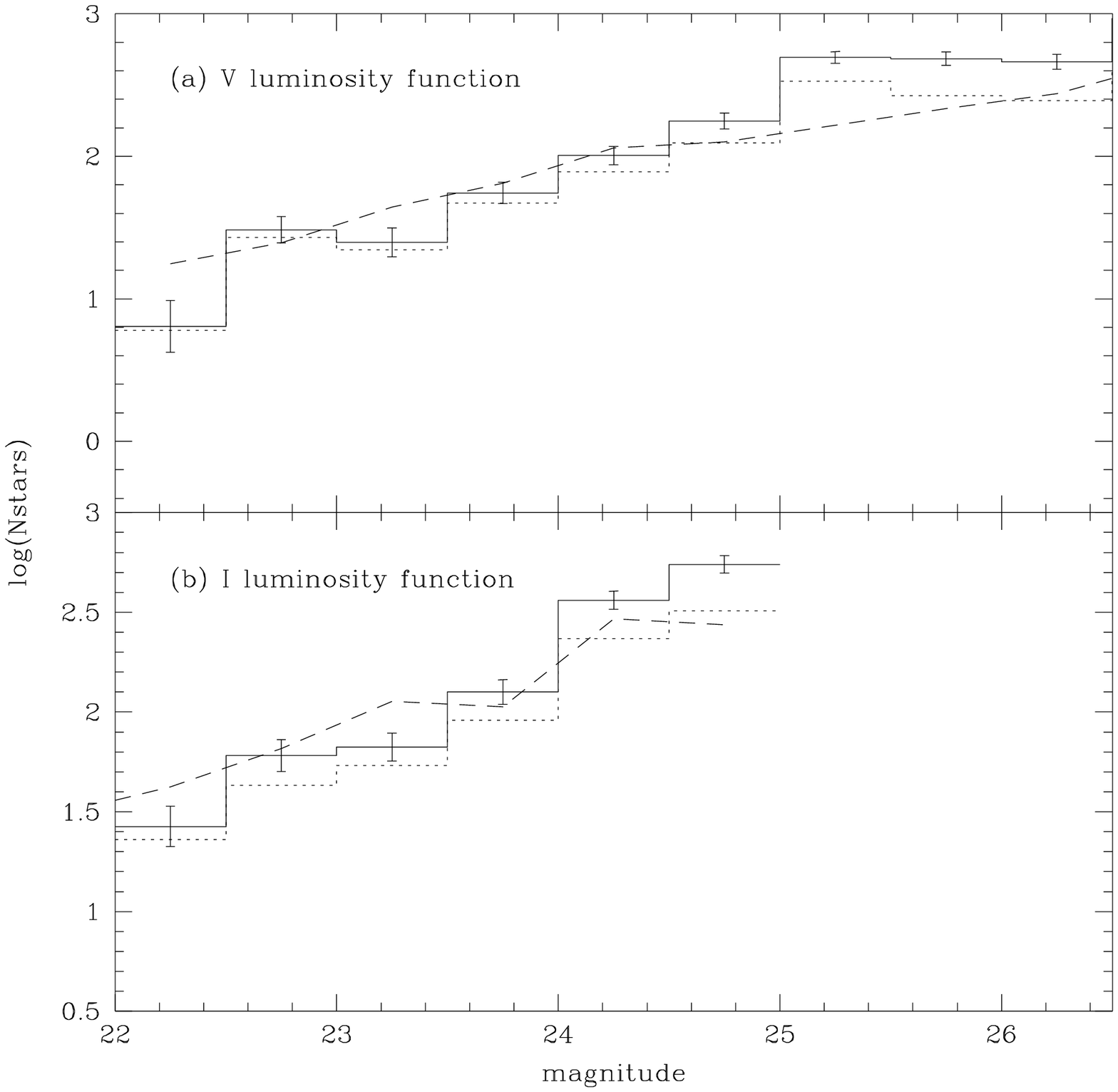]{Luminosity functions of the cluster in $V$ and $I$.  The dotted lines show the luminosity function before making completeness corrections; the solid lines and error bars are for the completed luminosity functions.  The dashed lines show the theoretical luminosity function, with horizontal branch stars omitted.\label{fig_lf}}
\figcaption[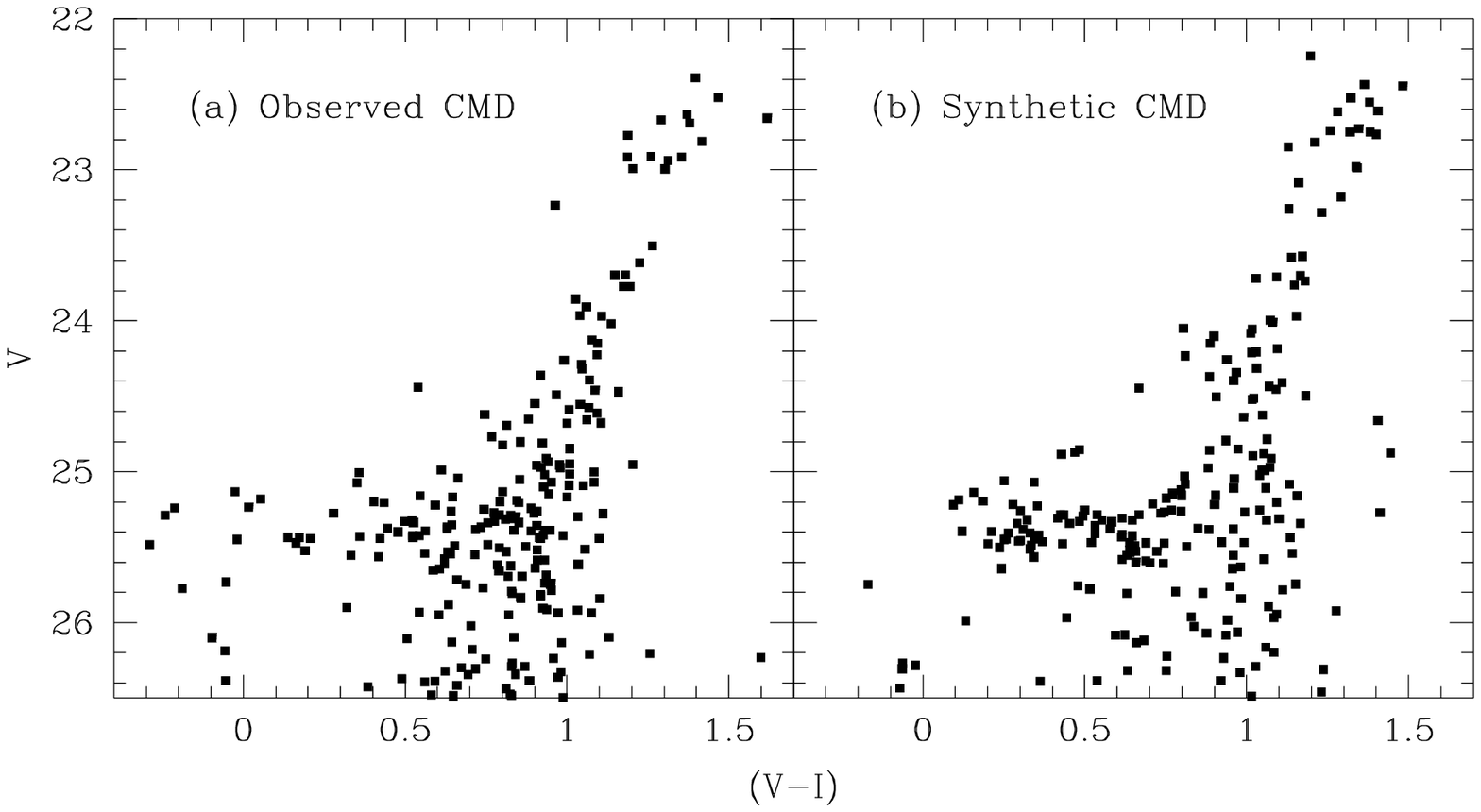]{The observed $V, V-I$ color-magnitude diagram of the cluster is shown in figure (a).  Figure (b) shows a synthetic CMD created using the solved star formation parameters\label{fig_cmd}}
\figcaption[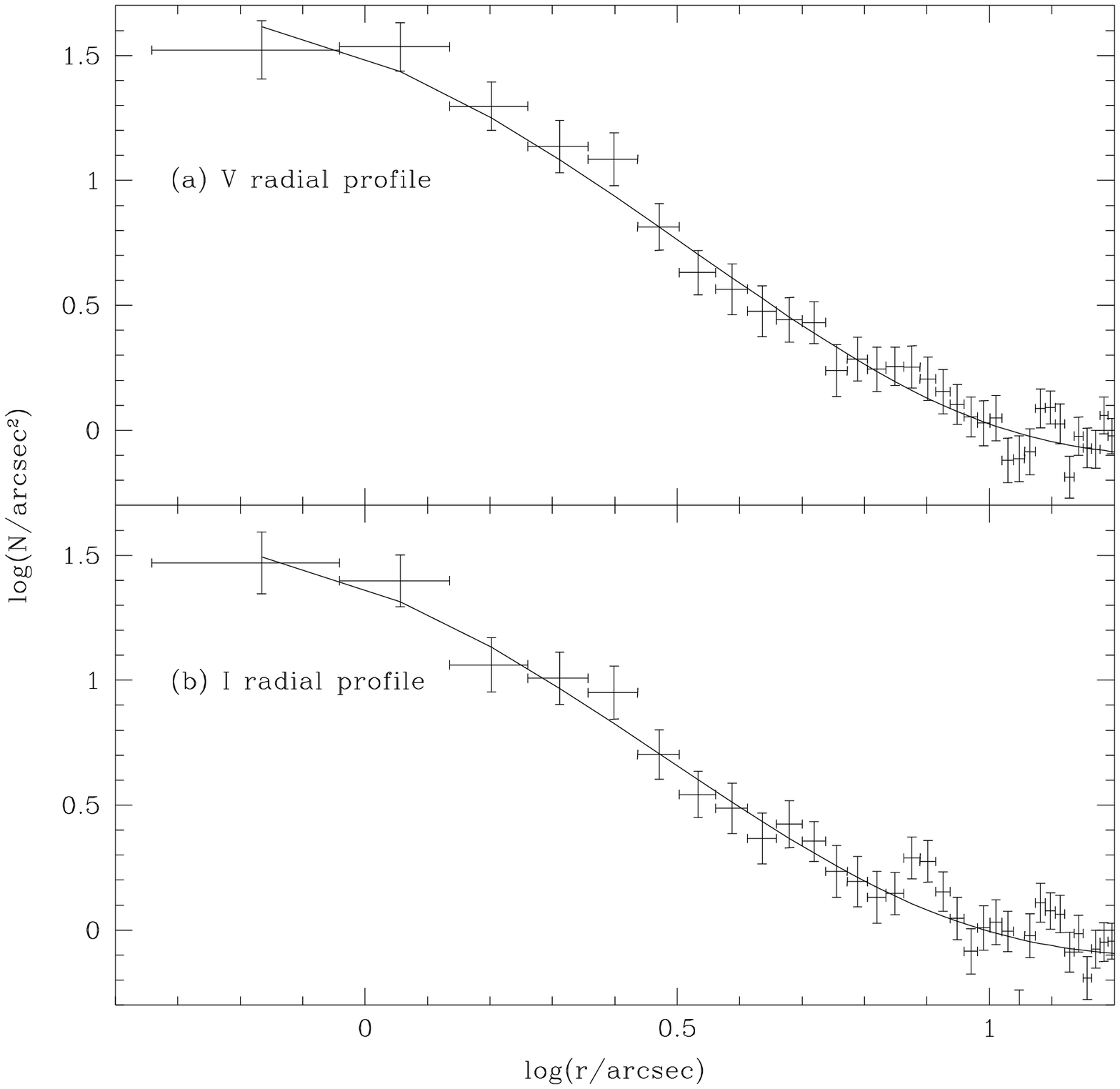]{Radial profile of the cluster in (a) $V$ for stars brighter than $V=27$ and (b) $I$ for stars brighter than $I=26$. The solid line is for a King model with r$_c$=1.14 arcsec and r$_t$=25 arcsec.\label{fig_prof}}

\end{document}